\documentclass[reprint,superscriptaddress,amsmath,amssymb,aps,pra,floatfix]{revtex4-2}

\usepackage{graphicx}
\usepackage{dcolumn}
\usepackage{bm}
\usepackage{url}
\usepackage[utf8]{inputenc}
\usepackage[T1]{fontenc}
\usepackage{booktabs, array, mathptmx, float, tabularx, booktabs, lipsum, amsmath,multirow}
\usepackage{siunitx, xcolor}
\usepackage[version=4]{mhchem}
\graphicspath{{figs/}{figsgaoerb/}}
\usepackage[colorlinks,linkcolor=blue,anchorcolor=blue,citecolor=blue,urlcolor=blue]{hyperref}

\begin{document}

\title{Electromagnetically driven, environmentally adaptive, and functionally switchable hydrodynamic devices}

\author{Chen-Long Wu}
\affiliation{School of Mechanical and Power Engineering, East China University of Science and Technology, Shanghai 200237, China}
\author{Bin Wang}\email{\textcolor{black}{bwang@ecust.edu.cn}}
\affiliation{School of Mechanical and Power Engineering, East China University of Science and Technology, Shanghai 200237, China}
\author{Hao Wang}
\affiliation{School of Mechanical and Power Engineering, East China University of Science and Technology, Shanghai 200237, China}
\author{Neng-Zhi Yao}
\affiliation{School of Mechanical and Power Engineering, East China University of Science and Technology, Shanghai 200237, China}
\author{Liujun Xu}
\affiliation{Graduate School of China Academy of Engineering Physics, Beijing 100193, China}
\author{Xuesheng Wang}\email{\textcolor{black}{wangxs@ecust.edu.cn}}
\affiliation{School of Mechanical and Power Engineering, East China University of Science and Technology, Shanghai 200237, China}
\author{Jiping Huang}\email{\textcolor{black}{jphuang@fudan.edu.cn}}
\affiliation{Department of Physics, State Key Laboratory of Surface Physics, and Key Laboratory of Micro and Nano Photonic Structures (MOE), Fudan University, Shanghai 200438, China}

\date{\today}

\begin{abstract}

Metamaterials provide exceptional control over physical phenomena, enabling many disruptive technologies. However, researches in hydrodynamic meta-devices have mainly used intrusive methods to manipulate material structures, limited by material properties and specific environmental conditions. Each design serves a single function, reducing versatility. This study introduces a meta-hydrodynamics theory using applied force fields to avoid physical contact with the fluid and eliminate the need for inhomogeneous and anisotropic metamaterials, allowing continuous switching between cloaking, shielding, and Venturi amplification. The force field operates independently of the fluid's physical properties, making it adaptable to various fluids and environmental conditions. We derive volumetric force distributions for hydrodynamic devices based on fluid properties and forces equivalence, using the integral median theorem to homogenize these forces for practical applications. The effectiveness of the proposed hydrodynamic devices is validated through numerical simulations and quantitative analyses. By utilizing the electromagnetic forces produced by the interaction between a conducting fluid and an electromagnetic field, we experimentally verified the validity of our theoretical simulations. Our research offers different insights into hydrodynamic meta-devices design, enhancing practical applications and opening avenues for innovative flow manipulation.

\end{abstract}

\maketitle

\section{Introduction}
Flow phenomena are pivotal in both natural and engineering applications, encompassing microscopic cell movements \cite{keren2009intracellular} to macroscopic oceanic and atmospheric cycles \cite{pedlosky2013geophysical}. These phenomena influence the stability of ecosystems and the survival of living organisms while also being crucial in fields such as energy, environmental protection, and industrial production. Therefore, understanding and manipulating flow phenomena are essential for advancing scientific progress and fostering technological innovation.

Recently, hydrodynamic metamaterials \cite{morton2008hydrodynamic} have garnered significant research attention due to their remarkable potential in manipulating fluid flows. Hydrodynamic metamaterials are an extension of metamaterials \cite{viktor1968electrodynamics, pendry1996extremely, pendry1999magnetism, shelby2001experimental} in the field of hydrodynamics, which are engineered with physical properties surpassing those of natural materials, have received extensive attention and research in systems such as wave and diffusion. These metamaterials manipulate various physical fields through the precise designs and combination of microstructures to achieve various functions such as cloaking, shielding, and amplification. In 2006, Pendry \textit{et al}. \cite{pendry2006controlling} demonstrate the formal invariance of Maxwell’s equations and introduce a theory of transformation optics based on coordinate transformations. This theory provides a foundational framework for the design of meta-devices such as cloaks, shields, and concentrators \cite{cai2007optical, valentine2009optical, ergin2010three}. Subsequently, this idea of equating material physical properties with spaces, based on the formal invariance of the control equations, is extended to other physical fields. These expansions facilitate the theoretical and experimental realizations of electromagnetic \cite{schurig2006metamaterial, li2008hiding, liu2009broadband}, acoustic wave \cite{cummer2008scattering, zhang2011broadband, popa2011experimental}, and thermal meta-devices \cite{fan2008shaped, li2015temperature, shen2016temperature, zhang2023diffusion, jin2023tunable, xu2023giant, yang2024controlling, fan2024thermal, wu2024optimal}.

The groundbreaking concept of hydrodynamic metamaterial cloaks are first introduced by Urzhumov \textit{et al}. \cite{urzhumov2011fluid, urzhumov2012flow}, revealing the potential of hydrodynamic meta-devices. This technology utilizes coordinate transformations in the flows through porous media, governed by Darcy’s law, to hide a three-dimensional sphere \cite{urzhumov2011fluid} and a two-dimensional cylinder \cite{urzhumov2012flow} by manipulating the permeability of the surrounding porous media. Follow-up experiments have successfully verified the effectiveness of this permeability manipulation \cite{chen2022realizing}. To lift the limitations imposed by porous media, Park \textit{et al}. \cite{park2019hydrodynamic, park2021metamaterial, park2019fluid} prove that the Stokes equations are formally invariant under viscous potential flows. They develop a dynamic viscosity tensor through coordinate transformations to realize hydrodynamic cloaks \cite{park2019hydrodynamic}, concentrators \cite{park2021metamaterial}, and rotators \cite{park2019fluid} in non-porous creeping flows by directing the viscous forces. Experimentally, the realization of this inhomogeneous and anisotropic dynamic viscosity tensor requires the placement of micropillars in the flow field. This process, however, inevitably results in an intrusion into the flow fields, and a single design can serve only one specific function. In pursuit of a simpler design, Tay \textit{et al}. \cite{tay2022metamaterial} achieve a hydrodynamic cloak without metamaterials. They alter the depth of the flow channel around the cylinder based on the scattering cancellation method. Beyond these passive techniques, active strategies involving applied external fields have also been explored for hydrodynamic cloaks. Wang \textit{et al}. \cite{wang2021intangible} design a hydrodynamic cloak with uniform dynamic viscosity using the convection-diffusion-balance method and modulate dynamic viscosity through an applied temperature field. Nonetheless, precisely controlling the viscosity of fluids remains challenging due to mass transport. Subsequently, Wu \textit{et al}. \cite{wu2024meta-hydrodynamics} establish an equivalence relation between volumetric forces and spaces, deriving the volumetric force distributions for hydrodynamic cloaks, concentrators, and rotators through coordinate transformations. However, these distributions are highly inhomogeneous and challenging to theoretically homogenize, posing significant difficulties for experimental realizations and practical applications. Moreover, Boyko \textit{et al}. \cite{boyko2021microscale} have pioneered a method involving the application of an electric field to generate electro-osmotic flows, adjusting velocity and pressure distributions around an obstacle to achieve hydrodynamic cloaking and shielding. This approach exploits the fact that the microscale flow fields are entirely dominated by boundary momentum sources and can be linearly superimposed, and therefore it is valid only at microcosmic scales.

Although several multifunctional hydrodynamic meta-devices have been proposed \cite{chen2022realizingpnas, jiang2025achieving}, they are primarily based on porous media flows by using intrusive micropillars. Realizing non-intrusive hydrodynamic meta-devices with multifunctionality in non-porous media flows remains a significant challenge. To address the challenges mentioned above, we propose a meta-hydrodynamics theory that employs external fields to achieve various functions. This theory allows continuous switching between functions such as cloaking, shielding, and amplification, without being restricted by the complex metamaterials and can be applied to different environmental conditions. We derive the exact analytical expressions for the required volumetric force distributions based on the equivalence between viscous and volumetric forces. To simplify the applications, we employ the integral median theorem to homogenize the volumetric force distributions, ensuring that the resulting force field retains components in only one direction. By leveraging the electromagnetic forces generated from the interaction between the conducting fluid and the electromagnetic field, we successfully achieve hydrodynamic cloaking, shielding, and amplification in the experiments. This hydrodynamic device successfully hides a two-dimensional cylindrical obstacle in the viscous potential flows under the cloaking function, and effectively prevents (or promotes) fluid entry into the central region under the shielding (or amplification) function. Our findings are expected to significantly advance the development of hydrodynamic meta-devices, thereby facilitating their practical applications in chemical, biological, and medical fields.
\begin{figure*}[!htb]
\setlength{\abovecaptionskip}{0cm}
\centering
\includegraphics[width=17.8cm]{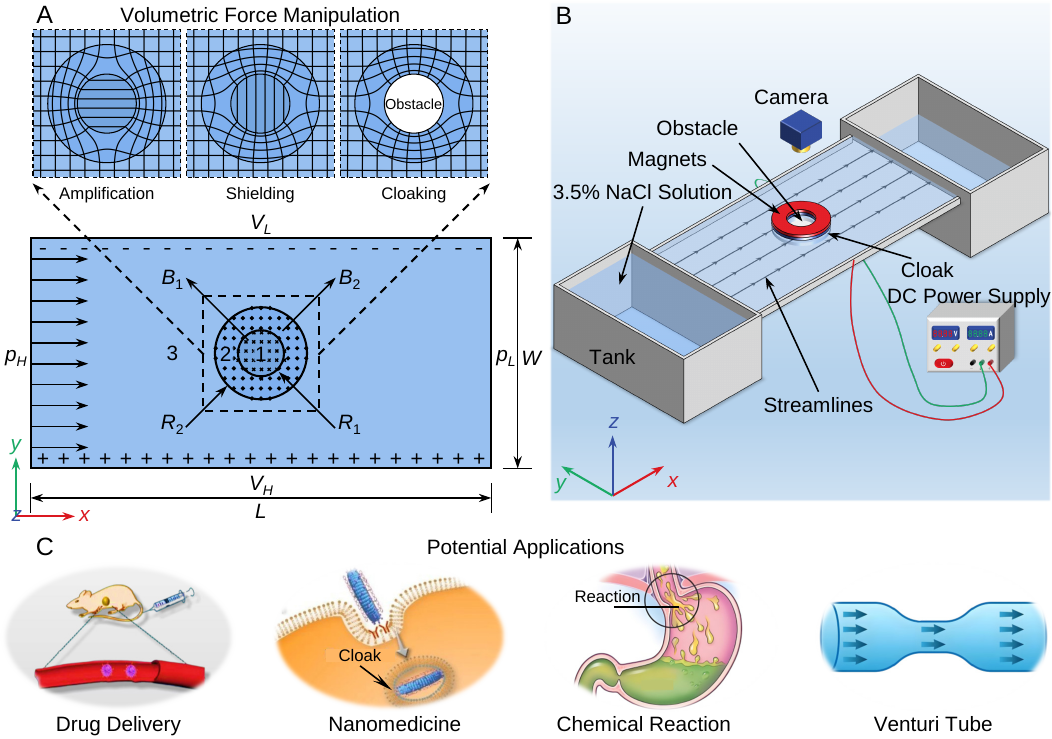}
\caption{Schematics of the hydrodynamic devices. (A) Schematic of the hydrodynamic devices with Venturi amplification, shielding and cloaking functions. The amplification (shielding) function promotes (prevents) fluid from entering the central region without disturbing the background flow field; the cloaking function removes interference with the flow field caused by obstacle in the center. The $x-y$ plane for the shielding function of the hydrodynamic devices, and the magnetic field in the opposite direction enables the amplification function. The hydrodynamic devices have inner and outer radii denoted as $R_1$ and $R_2$, respectively. In regions 1 and 2, the magnetic induction intensities are $B_1$ and $B_2$, respectively. (B) Schematic of the experimental setup for the cloaking function of the hydrodynamic devices, where the flow channel has the same three-dimensional dimensions as the model in (A). (C) The ability of hydrodynamic devices to function independently of the background environment suggests potential applications in areas such as drug delivery, nanomedicine, chemical reaction, and Venturi tube.}
\label{fig1}
\end{figure*}
\begin{figure*}[!htb]
\setlength{\abovecaptionskip}{0cm}
\centering
\includegraphics[width=17.8cm]{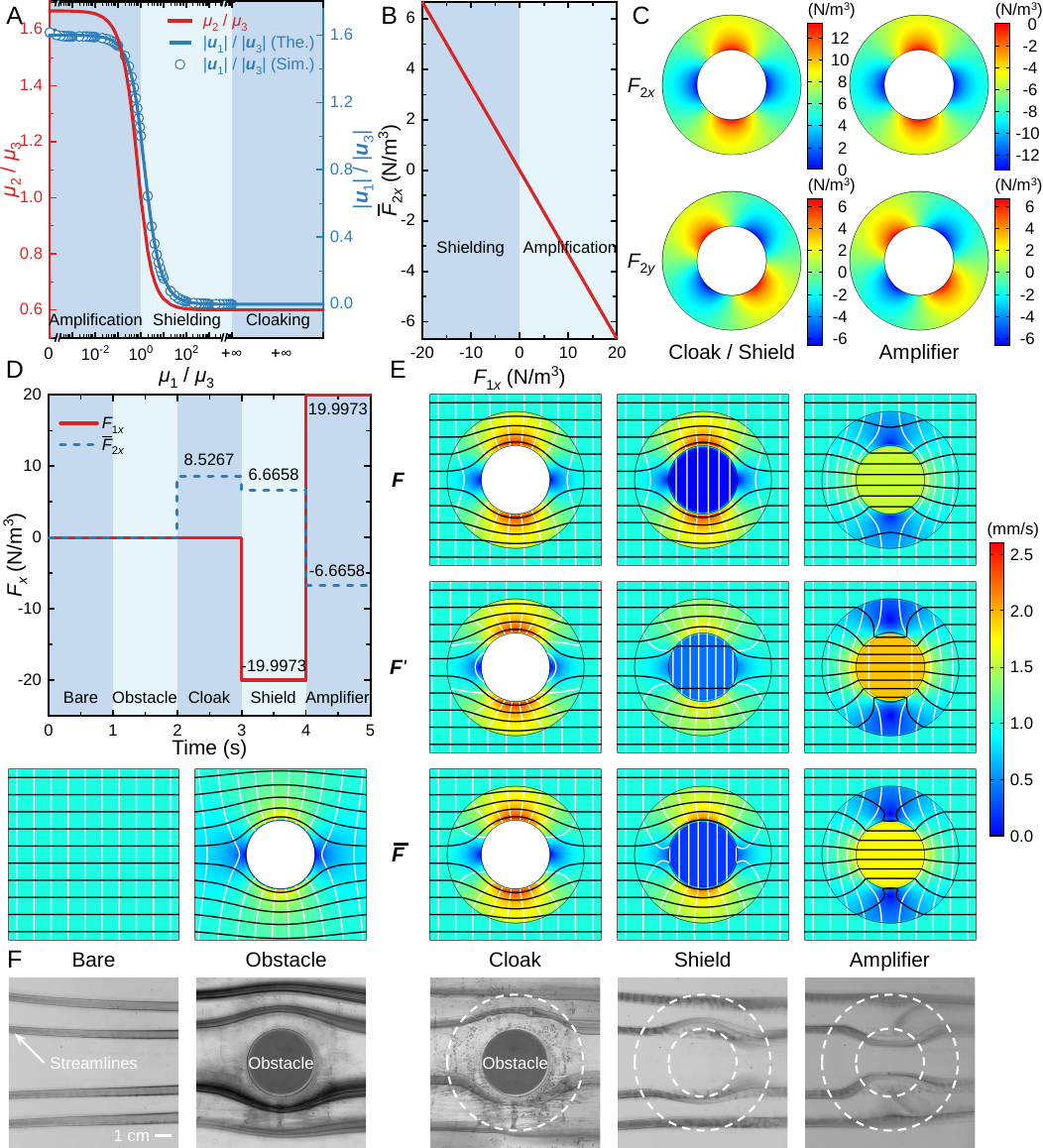}
\caption{Numerical simulation and experimental validation results. (A) Relationship between the manipulation effects of hydrodynamic devices and the hypothesized viscosity across various regions. (B) Linear relationship between ${\bar{F}}_{2x}$ and $F_{1x}$. (C) Volumetric forces $\boldsymbol{F}_2$ for cloaking, shielding, and amplification, where the negative sign indicates the opposite direction to the positive direction of the coordinate axis. (D) Characterization of the temporal dynamics of volumetric forces. (E) Velocity fields (localized) for the cases of the bare, obstacle, cloaking, shielding, and Venturi amplification, where the black lines represent streamlines and the white lines denote isobars. (F) Experimentally observed streamlines (black), where the fluid flows from the left to the right side of the photos. The white dotted line delineates the range of the hydrodynamic devices.}
\label{fig2}
\end{figure*}

\section{Results}
\subsection{Analytic theory}

Consider the flow model illustrated in \mbox{Fig. \ref{fig1}(A)}, planar flows of the incompressible viscous fluid with the dimensions of $L\times W=400\ \mathrm{mm}\times200\ \mathrm{mm}$. The model’s left boundary maintains a high-pressure of $p_H=5\ \mathrm{Pa}$, while the right boundary is set to a low-pressure of $p_L=0\ \mathrm{Pa}$. The upper and lower boundaries are set to the non-slip wall. This setup induces fluid flows driven by the pressure difference $\Delta p=p_H-p_L$ along the $x$-direction. The model delineates three different regions: the central function region ($r\le R_1=20\ \mathrm{mm}$), the hydrodynamic devices region ($R_1<r\le R_2$), and the background freestream region ($r>R_2=40\ \mathrm{mm}$).

For the steady-state incompressible flows depicted in \mbox{Fig. \ref{fig1}(A)}, the continuity and momentum equations can be expressed as \cite{batchelor1967introduction}
\begin{align}
\nabla\cdot\boldsymbol{u}=0.
\label{eq1}
\end{align}
\begin{align}
\rho\left(\boldsymbol{u}\cdot\nabla\right)\boldsymbol{u}=-\nabla p+\mu\nabla^2\boldsymbol{u}+\boldsymbol{F},
\label{eq2}
\end{align}
where $\rho$ and $\mu$ represent the density and dynamic viscosity of the fluids, respectively; $\boldsymbol{u}$ and $p$ denote the velocity vector and pressure, respectively; $\boldsymbol{F}$ signifies the volumetric forces acting on the fluids.

For viscous potential flows, when the volumetric forces are neglected, Eq. (\ref{eq2}) can be transformed into the Laplace’s equation. Assuming the fluid’s dynamic viscosity varies across different regions, we can derive a matching relation between viscosities and radii using the continuity conditions at the boundaries, as follows
\begin{align}
\left(\mu_1-\mu_2\right)\left(\mu_2+\mu_3\right)R_1^2+\left(\mu_1+\mu_2\right)\left(\mu_2-\mu_3\right)R_2^2=0.
\label{eq3}
\end{align}
This relation ensures minimal interference with the background flow field, and varying dynamic viscosity values lead to distinct manipulation effects, as illustrated in Fig.\ref{fig2}(A). Generally, the effectiveness of the hydrodynamic device in manipulating the flow field is governed by the ratio between the velocity in the central region and that in the background region. As illustrated in Fig. \ref{fig2}(A), when $\mu_1/\mu_3>1$, Eq. (\ref{eq3}) indicates that $\mu_2/\mu_3<1$, resulting in the velocity ratio of the velocity $\left|\boldsymbol{u}_1\right|$ in the center region to the velocity $\left|\boldsymbol{u}_3\right|$ in the background region is less than 1. This indicates that the fluid experiences greater resistance when entering the center region, thus representing the shielding case. Conversely, when $\mu_1/\mu_3<1$, $\mu_2/\mu_3>1$ and the velocity ratio $\left|\boldsymbol{u}_1\right|/\left|\boldsymbol{u}_3\right|$ exceeds 1, which characterizes the amplification case. The detailed derivation is in the SI Appendix, Section S1 \cite{supplementary}.

The above manipulation is achieved by adjusting the fluid's dynamic viscosity, but controlling viscosity precisely in each region is challenging due to mass transport complexities, especially when $\mu_1/\mu_3$ approaches 0 or infinity. There is an equivalence between fluid properties and volumetric forces, the detailed derivation is given in the SI Appendix, Section S2 \cite{supplementary}. This equivalence indicates that the effects of manipulating fluid properties can be replaced by applying volumetric forces, thereby eliminating the necessity to change the fluid’s viscosity. This allows us to derive the analytical expressions of volumetric forces for hydrodynamic devices as follows
\begin{align}
\boldsymbol{F}_1=\frac{1}{2}\left(1-\frac{\mu_3}{\mu_1}\right)\left(\begin{matrix}\left(1+\frac{\mu_2}{\mu_3}\right)+\left(1-\frac{\mu_2}{\mu_3}\right)\frac{R_2^2}{R_1^2}\\0\\\end{matrix}\right)\left(\frac{p_L-p_H}{L}\right),
\label{eq4}
\end{align}
\begin{align}
\boldsymbol{F}_2=\frac{1}{2}\left(1-\frac{\mu_3}{\mu_2}\right)\left(\begin{matrix}\left(1+\frac{\mu_2}{\mu_3}\right)-\left(1-\frac{\mu_2}{\mu_3}\right)R_2^2\frac{\cos{2\theta}}{r^2}\\-\left(1-\frac{\mu_2}{\mu_3}\right)R_2^2\frac{\sin{2\theta}}{r^2}\\\end{matrix}\right)\left(\frac{p_L-p_H}{L}\right).
\label{eq5}
\end{align}
To achieve the desired manipulation effect, we initially assume a value for $\mu_1/\mu_3$. Then, using Eq. (\ref{eq3}), we determine the value of $\mu_2/\mu_3$ and substitute them into Eqs. (\ref{eq4}) and (\ref{eq5}) to calculate the required volumetric forces. Significantly, the viscosity does not play a practical role in this process. When using volumetric forces for manipulation, it is no longer necessary to change the viscosity of areas 1 and 2. Each $\left|\boldsymbol{u}_1\right|/\left|\boldsymbol{u}_3\right|$ uniquely determines the corresponding pairs of $\mu_1/\mu_3$ and $\mu_2/\mu_3$, as shown in Fig. \ref{fig2}(A). Consequently, Eqs. (\ref{eq4}) and (\ref{eq5}) reveal that the identical volumetric forces produce the same manipulation effect ($\left|\boldsymbol{u}_1\right|/\left|\boldsymbol{u}_3\right|$) even for different fluid media (different $\mu_3$). This indicates that the force field operates independently of the fluid’s physical properties, allowing it to be applied across various fluid media, making it adaptable to different environmental conditions.

Due to the rotational symmetry in the geometry of the studied model [Fig. \ref{fig1}(A)], there exists another distributions for $\boldsymbol{F}_2$, which can be expressed as
\begin{align}
\boldsymbol{F}_2^\prime=\frac{1}{2}\left(1-\frac{\mu_3}{\mu_2}\right)\left(\begin{matrix}\left(1+\frac{\mu_2}{\mu_3}\right)+\left(1-\frac{\mu_2}{\mu_3}\right)R_2^2\frac{\cos{2\theta}}{r^2}\\\left(1-\frac{\mu_2}{\mu_3}\right)R_2^2\frac{\sin{2\theta}}{r^2}\\\end{matrix}\right)\left(\frac{p_L-p_H}{L}\right).
\label{eq6}
\end{align}

The volumetric forces, as delineated in Eqs. (\ref{eq5}) and (\ref{eq6}), exhibit the non-uniform distributions. This non-uniformity presents significant challenges for both experimental realizations and practical applications. To address this issue, we propose the utilization of the median of integral theorem to achieve homogenization. Consequently, the integral means of the volumetric forces are as follows
\begin{align}
\bar{\boldsymbol{F}_2}=\frac{\iint_{\mathnormal{\Omega}_2}\boldsymbol{F}_2d\Omega}{\iint_{\mathnormal{\Omega}_2} d\Omega}=\frac{1}{2}\left(1-\frac{\mu_3}{\mu_2}\right)\left(\begin{matrix}1+\frac{\mu_2}{\mu_3}\\0\\\end{matrix}\right)\left(\frac{p_L-p_H}{L}\right),
\label{eq7}
\end{align}
where $\mathnormal{\Omega}_2$ denotes the hydrodynamic devices region (region 2). As illustrated in Fig. \ref{fig2}(B), there is a linear relationship between ${\bar{F}}_{2x}$ and $F_{1x}$, demonstrating that ${\bar{F}}_{2x}=R_1^2/(R_1^2-R_2^2)F_{1x}$, where $F_{1x}$ ranges from $(2R_2^2/(R_2^2+R_1^2))((p_L-p_H)/L)$ to $(2R_2^2/(R_2^2+R_1^2)((p_H-p_L)/L)$.

This substitution of external forces for physical properties greatly simplifies the complexity of the problem of manipulating flows with mass transport. This treatment is effective in manipulating the fluid from a mechanics point of view alone. Hence, the method is tentatively referred to as the “meta-hydrodynamics \cite{wu2024meta-hydrodynamics}” .

Generally, there are several ways to obtain volumetric forces, including gravity, electromagnetic forces, inertial forces, and among others. Here, the discussion focuses on employing electromagnetic forces to achieve the specific volumetric force distributions. Under the assumption that, in the creeping flows, the induced electric and magnetic fields from fluid motion are negligible when compared to the externally applied electric and magnetic fields. This simplifies the problem to a one-way coupling issue where the fluid flows do not influence the electric and magnetic fields, but in turn the interaction of the electric and magnetic fields affects the fluid flows. We consider the arrangement of the electric and magnetic fields shown in Fig. \ref{fig1}(A), where the electric field aligns the positive direction of the $y$-axis, and the magnetic field does so along the $z$-axis. In the configuration of the electric field illustrated in Fig. \ref{fig1}(A), the bottom boundary is characterized by a high potential $V_H$ of $5\ \mathrm{V}$, while the top boundary corresponds to a low potential $V_L$ of $0\ \mathrm{V}$. Both the left and right boundaries are set to be electrically insulated. When an insulated solid obstacle is positioned in region 1, the electromagnetic force in region 2 can be derived as follows (SI Appendix, Section S4 \cite{supplementary})
\begin{align}
\boldsymbol{F}_{e,2}^\prime=\sigma\left(\frac{V_L-V_H}{W}\right)\left(\begin{matrix}-R_1^2\frac{\cos{2\theta}}{r^2}-1\\-R_1^2\frac{\sin{2\theta}}{r^2}\\\end{matrix}\right)B_2.
\label{eq8}
\end{align}
It is evident that the distribution of the electromagnetic force $\boldsymbol{F}_{e,2}^\prime$ mirrors that of the volumetric force $\boldsymbol{F}_2^\prime$ \mbox{(Eq. (\ref{eq6}))}. Therefore, this distribution can be utilized to achieve hydrodynamic cloaking. When a solid obstacle with the same conductivity as the fluid is positioned in the central region, or when the central region consists of fluid, the electromagnetic force exerted on the fluid is
\begin{align}
\boldsymbol{F}_{e,1}=\sigma\left(\frac{V_L-V_H}{W}\right)\left(\begin{matrix}-1\\0\\\end{matrix}\right)B_1,
\label{eq9}
\end{align}
\begin{align}
{\bar{\boldsymbol{F}}}_{e,2}=\sigma\left(\frac{V_L-V_H}{W}\right)\left(\begin{matrix}-1\\0\\\end{matrix}\right)B_2,
\label{eq10}
\end{align}
where $B_1$ (or $B_2$) is the magnetic induction intensity in region 1 (or 2). At this point, the electromagnetic forces are uniform, matching the volumetric force distributions described in \mbox{Eqs. (\ref{eq4}, \ref{eq7})}. This implies that dynamic switching between cloaking, shielding, and amplification can be accomplished using the same configuration.

\subsection{Simulation and experimental results}

To validate the effectiveness of the volumetric force distributions outlined in \mbox{Eqs. (\ref{eq4})-(\ref{eq7})}, the simulations are performed using the commercial finite element analysis software, COMSOL Multiphysics. To achieve viscous potential flows, we chose an experimental model $H=1\ \mathrm{mm}\ll L,W$, i.e., the Hele-Shaw flow \cite{hele1898flow}, as a validation model, but do not restrict to this model. Our simulations involve using a sodium chloride solution with a mass fraction of $3.5\%$, chosen for its salinity closely mirroring that of seawater. The physical properties of the solution are as follows: a density $\rho$ of $1024.75\ \mathrm{kg/m^3}$, a dynamic viscosity $\mu$ of $1.08\ \mathrm{mPa \cdot s}$, an electrical conductivity $\sigma$ of $4.788\ \mathrm{S/m}$, and a relative dielectric constant $\epsilon$ of $72$ \cite{hill2005sea, Maxwell1963sea, cox1970specific}. The freestream at an average velocity $U_\infty$ of $0.9645\ \mathrm{mm/s}$, yielding a Reynolds number of $\mathrm{Re}_{H,D}=\rho U_\infty H^2/\mu D=0.0228$. Significantly, the $H^2/D$ is regarded as the characteristic length for calculating the Reynolds number instead of $H$ \cite{tay2022metamaterial, wang2021intangible, dai2023transformation, yao2022convective}.

\begin{figure*}[!htb]
\setlength{\abovecaptionskip}{0cm}
\centering
\includegraphics[width=17.8cm]{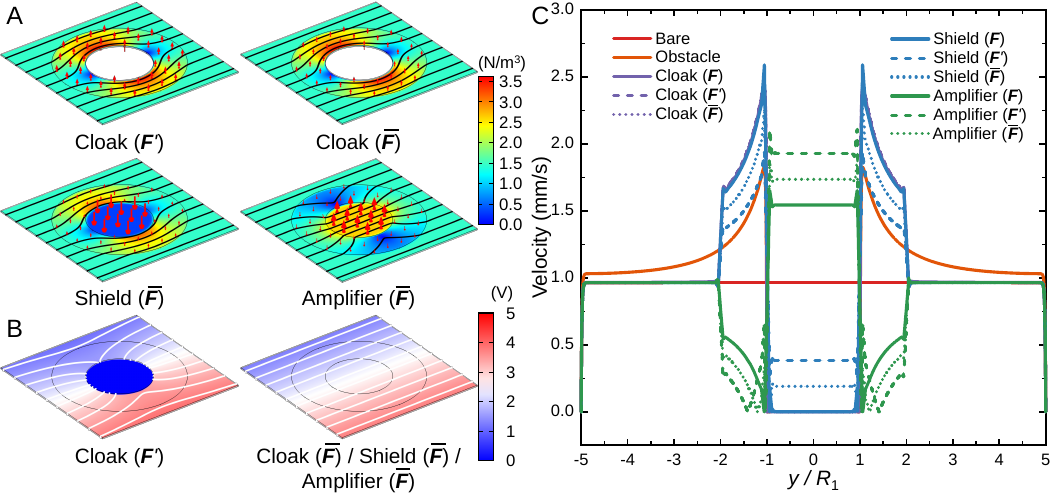}
\caption{Multiphysics simulation results, as well as comparisons of velocity distributions for hydrodynamic cloaks, shields, and amplifiers. (A, B) Multiphysics coupling results (plane of $z=0$, localized), where the black lines represent streamlines, the white lines indicate the equipotential lines, and the red arrows denote the direction of the magnetic field. (A) Cloaking ($\boldsymbol{F}^\prime$ and $\bar{\boldsymbol{F}}$), shielding ($\bar{\boldsymbol{F}}$), and amplification ($\bar{\boldsymbol{F}}$) under electromagnetic forces. (B) Potential distributions around an insulating obstacle and within pure fluids.(C) Velocity distributions along the characteristic line $x=0$ for different cases.}
\label{fig3}
\end{figure*}

Here, we consider two extreme cases: one for shielding, where $\mu_1/\mu_3 = 10^4$, and another for amplification, where $\mu_1/\mu_3 = 10^{-4}$. The distributions of $\boldsymbol{F}_2$ are illustrated in \mbox{Fig. \ref{fig2}(C)}, exhibiting obvious periodicity, directionality, and radial dependence. $\boldsymbol{F}_2^\prime$ can be considered the result of rotating $\boldsymbol{F}_2$ counterclockwise by $\pi/2$ in space, which are not shown here. 

The simulation results of the flow fields are presented in Fig. \ref{fig2}(E). Due to the significantly larger dimensions of the model in the $x$- and $y$-directions compared to the $z$-direction, the influence of the viscous boundary layers on the flow fields are minimal. Consequently, the velocity gradients, induced by the non-slip condition of the walls, are confined to the vicinity of the walls, ensuring a uniform flow velocity across the entire fields. This uniformity is evident in the perpendicular arrangement of the streamlines and isobars in the bare case. Introducing a cylindrical solid obstacle into the flow fields causes the fluid to divert around it, altering the trajectory of the streamlines. Meanwhile, the obstacle induces local variations in flow and pressure fields, characterized by a high-pressure region ahead of the obstacle and a low-pressure area behind it. To preserve mass conservation, the flow velocity increases on either side of the obstacle. When the volumetric forces are applied to the flow field, they prevent the fluid from entering the central region, effectively shielding it. Adjusting the direction of the volumetric forces can guide the fluid into the central region, thereby increasing the flow velocity there. These manipulations do not disturb the background flow field, which remains homogeneous with streamlines and isobars perpendicular to each other. Despite the varying distributions of these volumetric forces, they produce the same shielding or amplification effect. Remarkably, since a solid obstacle can be equated to an infinitely viscous fluid, the volumetric forces $\boldsymbol{F}_2$ used for shielding can also be employed to achieve cloaking over the obstacle. However, due to impedance mismatch at the boundaries, a compensation factor must be applied when using $\boldsymbol{F}_2^\prime$ or ${\bar{\boldsymbol{F}}}_2$ (SI Appendix, Section S3 \cite{supplementary}). These simulation results strongly validate our theoretical framework and demonstrate the effectiveness of the volumetric forces in manipulating fluid flows. Furthermore, as shown in \mbox{Fig. \ref{fig2}(D)}, these forces have a simple, adjustable form, allowing for continuous adjustment and dynamic switching for cloaking, shielding, and amplification (SI Appendix, Movie S1 \cite{supplementary}), demonstrates excellent transient performance. This result indicates that, while we derived analytical expressions for the volumetric forces using the steady-state control equations, these forces remain effective in transient states. The analytical expressions for the volumetric forces are derived based on the hypothesis of viscous potential flows. However, as flow velocity increases, the rotational effects in the flow field become evident. At this point, the flow field can no longer be considered viscous potential flows, leading to a gradual decline in the efficacy of the hydrodynamic devices. If the vortices in the flow field can be artificially eliminated, the proposed hydrodynamic devices would not be constrained by flow velocity (SI Appendix, Section S5 \cite{supplementary}).

As shown in Fig. \ref{fig1}(A), we construct a multiphysics model that couples electric, magnetic, and flow fields to assess the effectiveness of the electromagnetic forces. In our simulations, We adjust the electromagnetic forces by varying the magnetic induction intensities $B_1$ and $B_2$ in the direction along the $z$-axis direction. Our study indicates that the potential distributions around the insulating obstacle sufficiently meets the volumetric forces requirement specified in Eq. (\ref{eq7}). In this scenario, the magnetic induction $B_2$ in region 2 is $0.0949\ \mathrm{T}$. The simulation results of the multiphysics coupling are illustrated in Figs. \ref{fig3}(A, B), showing that an insulating obstacle causes equipotential lines to bend and increase in density around the obstacle, thereby enhancing the local electric field intensity [Fig. \ref{fig3}(B)]. The corresponding flow field is shown in Fig. \ref{fig3}(A), depicts how an non-uniform electric field interacts with a uniform magnetic field, creating the non-uniform electromagnetic forces that effectively hides the obstacle within the flow field. Additionally, when the conductivity of the solid matches that of the fluid, a consistent electric field intensity is observed. This uniformity is depicted by parallel and equally spaced equipotential lines, as shown in Fig. \ref{fig3}(B). Each equipotential line exhibits a consistent potential value, and the potential difference between adjacent lines remains uniform. This uniformity indicates that the potential changes linearly along the direction of the electric field. At this point, the magnetic induction $B_2$ in region 2 is $0.0712\ \mathrm{T}$. The associated flow field is displayed in Fig. \ref{fig3}(A), where a uniform electric field interacts with a magnetic field, generating the uniform electromagnetic forces. These forces accelerate the fluid flows, effectively enabling hydrodynamic cloaking by minimizing disturbances caused by obstacle. In cases of shielding or amplification, it is necessary to apply a magnetic field in both region 1 and region 2, as depicted in Fig. \ref{fig3}(A). For shielding, the magnetic fields B1 and B2 are $-0.1671\ \mathrm{T}$ and $0.0557\ \mathrm{T}$, respectively. Dynamic switching between shielding and amplification can be achieved simply by altering the direction of the magnetic fields. Although different conductive fluids have varying conductivities, which alter the magnitude of the electromagnetic forces. However, unlike conventional hydrodynamic meta-devices that require a redesign of the metamaterial structures for different fluids, our hydrodynamic devices only need an adjustment in the intensity of the electric or magnetic field to maintain optimal functionality. This flexibility allows our hydrodynamic devices to be highly adaptable to different environments and perform effectively under a wide range of conditions.

To quantitatively assess the performance of various hydrodynamic devices, we measure the velocity distributions along the central line ($x=0$) for each case. These measurements are compared against the control case (bare case), as presented in Fig. \ref{fig3}(C). In the background region, the flow velocities with the hydrodynamic devices mirrored those in the bare case, suggesting that the devices do not disturb the surrounding flow fields. In contrast, in the case without the devices (obstacle case), the flow velocities exhibited significant deviations from the bare case. Additionally, the effect of shielding (or amplification) on the flow field varies depending on the type of volumetric forces applied. As illustrated in Fig. \ref{fig3}(C), for shielding, the velocity in the central region under the influence of volumetric forces $\boldsymbol{F}$ is $0.0009\ \mathrm{mm/s}$. In contrast, the velocities in the central region under the influence of volumetric forces $\boldsymbol{F}^\prime$ and $\bar{\boldsymbol{F}}$ are $0.3849\ \mathrm{mm/s}$ and $0.1920\ \mathrm{mm/s}$, respectively. Similarly, for amplification, the velocities in the central region under the influence of volumetric forces $\boldsymbol{F}$, $\boldsymbol{F}^\prime$, and $\bar{\boldsymbol{F}}$ are $1.5441\ \mathrm{mm/s}$, $1.9299\ \mathrm{mm/s}$, and $1.7370\ \mathrm{mm/s}$, respectively. This feature belongs to the Venturi effect. However, unlike traditional applications of this effect, our hydrodynamic amplifiers achieve interference-free flow velocity amplification without the need to reduce the flow area.

To demonstrate and validate the designed hydrodynamic devices, we constructed an experimental setup as illustrated in Fig. \ref{fig1}(B). In the experiment, the Reynolds number ($\mathrm{Re}_{H,D}$) of the fluid flow is $\sim0.12$ (SI Appendix, Section S6 \cite{supplementary}). During the experiment, we quickly remove the top and bottom magnets to record the streamlines at this point, as presented in Fig. \ref{fig2}(F) (SI Appendix, Movie S2 \cite{supplementary}). The experimental results illustrate that the streamlines in the flow field are parallel to each other and nearly straight in the bare case. When a cylindrical glass obstacle is placed in the center of the flow field, the streamlines are deflected due to the obstacle obstructing the fluid flows and the fluid automatically bypasses the obstacle, resulting in the curved streamlines. However, when the obstacle is wrapped in a hydrodynamic cloak, the deflection of the streamlines outside the cloak is eliminated, resulting in straight streamlines that effectively hide the obstacle in the flow fields. When the central obstacle is removed and a magnetic field is introduced in its place, the fluid intuitively diverts away from the central region, thereby creating an effective shielding mechanism. Conversely, by adjusting the magnetic field's direction, the fluid can be guided back to converge at the center, resulting in flow amplification. Notably, the streamlines outside the hydrodynamic device remain straight and unaffected, regardless of whether cloaking, shielding, or amplification is in effect. Generally, the experimental results [Fig. \ref{fig2}(F)] and the simulation results [Fig. \ref{fig2}(E)] match each other, indicating the accuracy and reliability of our simulation processes. The experimental results effectively demonstrate the ability of volumetric forces to manipulate fluid flows. This manipulation capability is not only limited to hydrodynamic cloaking, shielding, and amplification, but can also be extended to other functions. This discovery implies that once people have mastered the physical mechanisms of the desired hydrodynamic devices, they can achieve multiple effects in different environments and applications by adjusting and optimizing the corresponding volumetric forces designs. This methodology is instrumental in advancing the fluid dynamics field and possesses the capability to spark innovation across various disciplines of physics and engineering.

\section{Discussion and Conclusion}

Overall, both experimental and simulation results validate the effectiveness of the proposed technique for achieving multiple hydrodynamic functions via external forces. These hydrodynamic devices hold significant potential applications in the biological, medical, chemical, and microfluidic \mbox{\cite{link2004geometrically, guo2012droplet, mazutis2013single}} fields [Fig. \ref{fig1}(C)]. Because some devices in these fields are typically used to manipulate and control tiny volumes of fluids, and flow channels in these devices are extremely narrow, the flow characteristics resemble Hele-Shaw flow. Therefore, the model we researched can help researchers efficiently control and mix minute quantities of fluids, with applications in biological sample analysis, drug delivery, and chemical reactions. For example, due to their ability to achieve zero interference with the surrounding environment, hydrodynamic cloaks can be instrumental in designing drug delivery systems. These devices can precisely target diseased areas, administering medication with minimal disruption to the surrounding environment. Additionally, the hydrodynamic shields protect the region containing normal cells from damage induced by the drug. Meanwhile, the hydrodynamic amplifiers enhance the concentration of the drug in areas with cancer cells, facilitating a more rapid destruction of the cancerous cells. Furthermore, hydrodynamic amplifiers can be utilized in cell culture, optimizing cell growth conditions by precisely controlling fluid flow direction and intensity without disturbing the background environment. In addition, hydrodynamic amplifiers can also enhance mixing and mass transfer in chemical reactors, thus improving the reaction efficiency and product quality. Moreover, hydrodynamic amplifiers can be applied to the design of novel interference-free Venturi tubes and interference-free Venturi meters.

In this study, we propose a different strategy to develop hydrodynamic devices capable of cloaking, shielding, and amplification using an externally applied force field. This approach is independent of metamaterials, requires no changes to the physical properties of the fluid, and applies to various fluid media. It is characterized by high flexibility and controllability, with the capability to be activated or deactivated as required and enabling continuous switching between various functions. We replace the effect of viscous forces with volumetric forces based on their equivalence relation and then derive the distributions of these forces. Employing the integral median theorem, these non-uniform volumetric forces are homogenized. This simplification reduces the forces to a single directional component and significantly simplifies their practical applications. Subsequently, the effectiveness of the volumetric forces is validated through numerical simulations and quantitative analyses. Experimentally, hydrodynamic cloaking, shielding, and amplification are achieved using the electromagnetic forces generated by the interaction between a conducting fluid and an electromagnetic field. The experiments confirm that the designed hydrodynamic device can effectively conceal a cylindrical obstacle in the viscous potential flows, in addition to achieving shielding and amplification of the flow field. Additionally, our approach is versatile, extending beyond circular-cylinder to encompass confocal elliptical-cylinder \cite{wu2024optimal} and arbitrary geometries \cite{xu2024free} as well as transient situations. In conclusion, the proposed theoretical framework offers novel perspectives for the development of hydrodynamic meta-devices, such as hydrodynamic  homogenizers \cite{chen2024rapidly}. Meanwhile, it can be applied to the design of other physical field meta-devices, including thermal \cite{guo2018thermal}, electric \cite{ma2013experiments}, acoustic \cite{shen2012acoustic}, and electromagnetic \cite{selvanayagam2013experimental} fields.

\section{Materials and Methods}

\subsection{Experimental demonstration}

The experiment involves the sodium chloride solution with a mass fraction of $3.5\%$ flowing in a shallow channel, formed by two thin glass plates of $1\ \mathrm{mm}$ thickness and separated by a gap of $1\ \mathrm{mm}$. The flow channel’s side walls are sealed by thin iron sheets of $1\ \mathrm{mm}$ thickness, and connected to the positive and negative terminals of a DC power supply, respectively. The ends of the flow channel are connected to two tanks, into which the solution is uniformly and continuously pumped via a peristaltic pump to maintain a constant liquid level. To visualize the flows, we inject the solution stained with black pigment into the flow channel using an injection pump connected to four symmetrically placed injection tubes. To demonstrate the cloaking function of the hydrodynamic devices, a cylindrical glass obstacle with a diameter of $40\ \mathrm{mm}$ and a height of $1\ \mathrm{mm}$ is placed in the middle of the channel. Toroidal ferrite magnets, featuring with an inner diameter of $40\ \mathrm{mm}$ and an outer diameter of $80\ \mathrm{mm}$ are positioned above and below the glass plates to create a vertically upward magnetic field. To verify the shielding function, maintain the toroidal magnets stationary, remove the central obstacle, and place the cylindrical magnets with a diameter of $40\ \mathrm{mm}$ above and below its position to generate a vertically downward magnetic field. Conversely, to validate the amplifying function, change the orientations of the toroidal and cylindrical magnets. The entire experimental process is captured on video by a camera positioned above the flow channel.

\bibliography{mybibfile}

\end{document}